\documentclass[12pt, draftclsnofoot, onecolumn]{IEEEtran}

\usepackage{amsmath,amssymb,amsfonts,amsthm}
\usepackage{algorithm,algorithmic}
\usepackage{xcolor}
\usepackage{stfloats,graphicx,romannum}
\usepackage[caption=false]{subfig}
\usepackage{enumitem}
\usepackage{balance}
\def\ie{{\it i.e.,\ \/}}
\def\st{{\it s.t.,\ \/}}
\def\eg{{\it e.g.,\ \/}}

\theoremstyle{definition}
\newtheorem{theorem}{Theorem}
\newtheorem{lemma}[theorem]{Lemma}
\newtheorem{corollary}[theorem]{Corollary}
\newtheorem{remark}{Remark}
\newtheorem{assumption}{Assumption}
\makeatletter
\def\blfootnote{\gdef\@thefnmark{}\@footnotetext}
\makeatother
\allowdisplaybreaks
\begin{document}

\setlength{\textfloatsep}{.75cm}

\pagenumbering{arabic}

%%%%%%%%%%%%%%%%%%%%%%%%%%%%%%%%%%%%%%%%%%%%%%%%%%%%%%%%%%%%
%                        Title                             %
%%%%%%%%%%%%%%%%%%%%%%%%%%%%%%%%%%%%%%%%%%%%%%%%%%%%%%%%%%%%

\title{Hierarchical Online Convex Optimization}
\author{ Juncheng Wang, Ben Liang, Min Dong, Gary Boudreau, and Hatem Abou-zeid
\thanks{J. Wang and B. Liang are with the University of Toronto (e-mail:
\{jcwang, liang\}@ece.utoronto.ca). M. Dong is with the Ontario Tech
University (e-mail: min.dong@ontariotechu.ca). G. Boudreau and H. Abou-zeid
are with Ericsson Canada (e-mail: \{gary.boudreau, hatem.abou-zeid\}@ericsson.com).}}
\maketitle

%%%%%%%%%%%%%%%%%%%%%%%%%%%%%%%%%%%%%%%%%%%%%%%%%%%%%%%%%%%%
%                       Abstract                           %
%%%%%%%%%%%%%%%%%%%%%%%%%%%%%%%%%%%%%%%%%%%%%%%%%%%%%%%%%%%%

\begin{abstract}
We consider online convex optimization (OCO) over a heterogeneous network with communication delay, where multiple workers together with a master execute a sequence of decisions to minimize the accumulation of time-varying global costs. The local data may not be independent or identically distributed, and the global cost functions may not be locally separable. Due to communication delay, neither the master nor the workers have in-time information about the current global cost function. We propose a new algorithm, termed Hierarchical OCO (HiOCO), which takes full advantage of the network heterogeneity in information timeliness and computation capacity to enable multi-step gradient descent at both the workers and the master. We analyze the impacts of the unique hierarchical architecture, multi-slot delay, and gradient estimation error to derive upper bounds on the dynamic regret of HiOCO, which measures the gap of costs between HiOCO and an offline globally optimal performance benchmark. 
\end{abstract}

%%%%%%%%%%%%%%%%%%%%%%%%%%%%%%%%%%%%%%%%%%%%%%%%%%%%%%%%%%%%
%                    Introduction                          %
%%%%%%%%%%%%%%%%%%%%%%%%%%%%%%%%%%%%%%%%%%%%%%%%%%%%%%%%%%%%

\section{Introduction}
\label{sec:Introduction}

Many machine learning, signal processing, and resource allocation problems can be cast into a dynamic optimization problem with time-varying convex cost functions. Online convex optimization (OCO) provides the tools to handle these dynamic problems in the presence of uncertainty, where an online decision strategy evolves based on the historical information~\cite{OLBK}. OCO can be seen as a discrete-time sequential learning and decision-making process by an agent in a system. At the beginning of each time slot, the agent makes a decision from a convex feasible set. The system reveals information about the current convex cost function to the agent only at the end of each time slot. The lack of in-time information prevents the agent from making an optimal decision at each time slot. Instead, the agent resorts to minimizing the \textit{regret}, which is the performance gap between the online decision sequence and some benchmark solution.

Most of the early works on OCO studied the \textit{static regret}, which compares the online decision sequence with a static offline benchmark \cite{Zinkevich03}\nocite{OT}\nocite{Slow}-\cite{AD}.
However, the optimum of dynamic problems is often time varying. As a rather coarse performance metric, achieving sublinear static regret may not be meaningful
since the static offline benchmark itself may perform poorly. The more attractive notion of \textit{dynamic regret} was also proposed in \cite{Zinkevich03}, where the offline benchmark solution can be time varying. Dynamic regret bounds are often expressed in terms of certain variation measures that reflect how dynamic the system is. Theoretical guarantees on the dynamic regret for convex and strongly convex cost functions were studied in \cite{Zinkevich03},~\cite{Hall15}-\cite{Jadbabaie15} and \cite{Mokhtari16}\nocite{Zhang17}\nocite{Bedi18}-\cite{Dixit19}, respectively.

The works \cite{Zinkevich03}\nocite{OT}\nocite{Slow}\nocite{AD}\nocite{Hall15}\nocite{Jadbabaie15}\nocite{Mokhtari16}\nocite{Zhang17}\nocite{Bedi18}-\cite{Dixit19} focused on centralized OCO, so they did not consider the network heterogeneity in information timeliness and computation capacity in many practical applications. For example, in mobile edge computing~\cite{LiangMEC17}, the local processors have timely information about their own computing tasks but may offload some tasks to the edge server due to the limitation on local computation resources. Another example is self-driving vehicular networks, where each vehicle moves based on its real-time sensor data while reporting local observations to a control center for traffic routing or utility maximization. In these examples, data are distributed away from the central coordinator and vary over time. Furthermore, the local devices need to make real-time decisions for global cost minimization. Applying centralized OCO approaches~\cite{Zinkevich03}\nocite{OT}\nocite{Slow}\nocite{AD}\nocite{Hall15}\nocite{Jadbabaie15}\nocite{Mokhtari16}\nocite{Zhang17}\nocite{Bedi18}-\cite{Dixit19} to these systems would lead to degraded performance, since they do not fully utilize the computing capability or information timeliness of all network nodes.

Existing works on distributed OCO \cite{Hosseini13}\nocite{Yan13}\nocite{Nedic15}\nocite{Tsianos16}\nocite{Ferdinand19}\nocite{Nunez14}\nocite{Akbari17}\nocite{Jadbabaie18}\nocite{Nima20}-\cite{Zhang19} are confined to \textit{separable} global cost functions, \ie they can be expressed as a sum of local cost functions. Specifically, each local cost function depends only on the local data, which allows each node to locally compute the gradient of its own cost function without information about the data at the other nodes. However, in many practical systems, including the aforementioned mobile edge computing and self-driving vehicular networks, the global cost functions are often \textit{non-separable}, due to the coupling of data or decision variables at different nodes. The solutions of \cite{Hosseini13}\nocite{Yan13}\nocite{Nedic15}\nocite{Tsianos16}\nocite{Ferdinand19}\nocite{Nunez14}\nocite{Akbari17}\nocite{Jadbabaie18}\nocite{Nima20}-\cite{Zhang19} are not applicable to such systems.

Furthermore, in practical systems, the decision makers often gain access to the system information only after some delay. In the standard OCO setting, the decision maker receives information about the current cost function at the end of each time slot when the decision is made, \ie the feedback information is delayed for only one time slot. This is the setting used in all existing distributed OCO works \cite{Hosseini13}\nocite{Yan13}\nocite{Nedic15}\nocite{Tsianos16}\nocite{Ferdinand19}\nocite{Nunez14}\nocite{Akbari17}\nocite{Jadbabaie18}\nocite{Nima20}-\cite{Zhang19}, but it can be too restrictive for many practical applications.

In this work, we aim to develop an online learning algorithm that takes full advantage of the network heterogeneity in information timeliness and computation capacity, while allowing the global cost functions to be non-separable and accommodating multi-slot information delay. The main contributions of this paper are as follows:

\begin{itemize}

\item We formulate a new OCO problem for a heterogenous master-worker network with communication delay, where the worker nodes have timely information about the local data but possibly less computation resources compared with the master node. At the beginning of each time slot, each worker node executes a local decision to minimize the accumulation of time-varying global costs. The local data at the worker nodes may not be independent or identically distributed, the global cost functions may be non-separable, and the feedback and data acquisition delay may span multiple time slots. This new problem formulation has many practical applications and broadens the scope of OCO in the existing literature.

\item We propose a new Hierarchical Online Convex Optimization (HiOCO) algorithm that takes full advantage of the network heterogeneity in information timeliness and computation capacity. HiOCO allows both \textit{timely local} gradient descent at the worker nodes and \textit{delayed global} gradient descent at the master node to improve system performance. Furthermore, HiOCO allows multi-step estimated gradient descent at both the worker nodes and the master node to fully utilize their computation resources. 

\item We analyze the special structure of HiOCO in terms of its hierarchical multi-step estimated gradient descent, in the presence of multi-slot delay. We prove that HiOCO yields $\mathcal{O}(\min\{\max\{\tau\Pi_T^\star,\Delta_{T}\},\max\{\tau^2\Pi_{2,T}^\star,\Delta_{2,T}\}\})$ dynamic regret, where $\tau$ is the total feedback delay, and $\Pi_T^\star$, $\Pi_{2,T}^\star$, $\Delta_T$, and $\Delta_{2,T}$ are certain variation measures that represent how dynamic the system is (see definitions in Section \ref{Sec:Metric}). We further extend our analysis to the case where even the worker nodes experience multi-slot delay to collect their local data.

\end{itemize}

The rest of this paper is organized as follows. In Section \ref{Sec:Related Work}, we present the related work. Section \ref{Sec:Problem} describes the system model, problem formulation, and performance metrics for OCO. We present the HiOCO algorithm, prove its dynamic regret bounds, and discuss its performance merits in Section~\ref{Sec:HiOCO}. We further extend HiOCO to the case of non-zero local delay in Section \ref{sec:Extension with Local Delay}. Concluding remarks are provided in Section \ref{Sec:Conclusions}.

%%%%%%%%%%%%%%%%%%%%%%%%%%%%%%%%%%%%%%%%%%%%%%%%%%%%%%%%%%%%
%                    Related Work                          %
%%%%%%%%%%%%%%%%%%%%%%%%%%%%%%%%%%%%%%%%%%%%%%%%%%%%%%%%%%%%

\section{Related Work}
\label{Sec:Related Work}

In this section, we survey existing works on OCO.\footnote{We focus on common short-term constrained OCO. OCO with long-term constraints are out of the scope of this paper. We refer interested readers to \cite{Trade}\nocite{Yu-SC}-\cite{T.Chen} and references therein.} The differences between the existing literature and our work are summarized in Table \ref{Tab:RE}.

\subsection{Online Learning and OCO}

Online learning is a method of machine learning where a learner attempts to tackle some decision-making task by learning from a sequence of data instances.
In the seminal work of OCO \cite{Zinkevich03}, an online projected gradient descent algorithm achieves $\mathcal{O}(\sqrt{T})$ static regret with bounded feasible set and gradient, where $T$ is the time horizon. The static regret was shown to be unavoidably $\Omega(\sqrt{T})$ for general convex cost functions without additional assumptions, but it was further improved
to $\mathcal{O}(\log{T})$ for strongly convex cost functions~\cite{OT}. In the standard OCO setting, information feedback is delayed for only one time slot, which is restrictive for many practical applications. The standard online projected gradient descent algorithm was extended in \cite{Slow} to provide $\mathcal{O}(\sqrt{\tau{T}})$ static regret in the presence of $\tau$-slot delay. Moreover, \cite{AD} studied OCO with adversarial delay.

The dynamic regret of OCO was first introduced in \cite{Zinkevich03}, and it has received a recent surge of interest \cite{Hall15}, \cite{Jadbabaie15}. Strong convexity was shown to improve the dynamic regret bound in~\cite{Mokhtari16}. By increasing the number of gradient descent steps, the dynamic regret bound was further improved in \cite{Zhang17}. Furthermore, extensions to accommodate inexact gradient were proposed in \cite{Bedi18} and \cite{Dixit19} with dynamic regret analysis.

Centralized OCO algorithms \cite{Zinkevich03}\nocite{OT}\nocite{Slow}\nocite{AD}\nocite{Hall15}\nocite{Jadbabaie15}\nocite{Mokhtari16}\nocite{Zhang17}\nocite{Bedi18}-\cite{Dixit19} naturally do not require the global cost function to be separable. Therefore, they can be applied to the online optimization problem we consider in this work. However, this way of solving the problem does not utilize the information timeliness or computation capacity at all network nodes. This can lead to substantial degradation of system performance compared with the proposed HiOCO approach. 

\subsection{Distributed OCO}

\begin{table*}[!t]
\renewcommand{\arraystretch}{1.2}
\vspace{0mm}
\caption{Summary of Related Dynamic Regret Bounds for OCO}
\label{Tab:RE}
\vspace{-2mm}
\centering
\resizebox{\textwidth}{!}{
\begin{tabular}{|c|c|c|c|c|c|c|c|c|c|c|c|c|c|c|c|c|}\hline
References&Gradient calculation&Cost function&Non-separable cost&Inaccurate~gradient&Feedback~delay&Dynamic~regret~(see~definitions~in~Section~\ref{Sec:Metric})\\\hline\hline
\cite{Zinkevich03}               & Central &Convex &Allowed &Not allowed&$\tau=1$&$\mathcal{O}(\Pi_T\sqrt{T})$\\\hline
\cite{Jadbabaie15}               & Central &Convex &Allowed &Not allowed&$\tau=1$&$\mathcal{O}(\min\{\sqrt{\Gamma_{2,T}\Pi_T^\star},(\Gamma_{2,T}\Theta_TT)^{\frac{1}{3}}\})$\\\hline
\cite{Mokhtari16}                & Central &Strongly convex &Allowed &Not~allowed&$\tau=1$&$\mathcal{O}(\Pi_T^\star)$\\\hline
\cite{Zhang17}                   & Central &Strongly Convex &Allowed &Not~allowed&$\tau=1$&$\mathcal{O}(\min\{\Pi_T^\star,\Pi_{2,T}^\star\})$\\\hline
\cite{Bedi18}, \cite{Dixit19}    & Central &Strongly Convex &Allowed &Allowed&$\tau=1$&$\mathcal{O}(\max\{\Pi_T^\star,\Delta_T\})$\\\hline
\cite{Jadbabaie18}, \cite{Nima20}& Local   &Convex &Not allowed &Not allowed&$\tau=1$&$\mathcal{O}(\sqrt{\Pi_T^\star{T}})$\\\hline
\cite{Zhang19}                   & Local   &Strongly Convex &Not allowed&Not~allowed&$\tau=1$&$\mathcal{O}(\max\{\Pi_T^\star,\Gamma_T\})$\\\hline
HiOCO                  & Local and Central &Strongly Convex &Allowed &Allowed&$\tau\ge1$&$\mathcal{O}(\min\{\max\{\tau\Pi_T^\star,\Delta_{T}\},\max\{\tau^2\Pi_{2,T}^\star,\Delta_{2,T}\}\})$\\\hline
\end{tabular}
}
\vspace{0mm}
\end{table*}

Early works on distributed OCO focused on the static regret \cite{Hosseini13}\nocite{Yan13}\nocite{Nedic15}\nocite{Tsianos16}\nocite{Ferdinand19}\nocite{Nunez14}-\cite{Akbari17}. A distributed online algorithm based on dual averaging was proposed in \cite{Hosseini13}. It achieves $\mathcal{O}(\sqrt{T})$ static regret. Another dual averaging based algorithm was shown to yield $\mathcal{O}(\log{T})$ static regret for strongly convex  cost functions \cite{Yan13}. Further extensions include Nesterov's primal-dual approach~\cite{Nedic15} and approximate dual averaging~\cite{Tsianos16}. Any-bach dual averaging was proposed in~\cite{Ferdinand19}. The impact of network topology on the performance of distributed OCO was studied in \cite{Nunez14} and~\cite{Akbari17}.

More recent works on distributed OCO considered the dynamic regret, since it is a better measurement of performance when the system environment is time-varying. Distributed mirror descent for online optimization was studied in \cite{Jadbabaie18}. Any-batch mirror descent was proposed in~\cite{Nima20} for heterogeneous networks. In \cite{Zhang19}, a push-pull based online algorithm provided an improved dynamic regret for strongly convex cost functions.

Existing works on distributed OCO are confined to separable global cost functions. For non-separable global cost  minimization in distributed networks,  more information exchange is needed among the network nodes. Furthermore, these works are under the standard OCO setting with one-slot feedback delay. In contrast, HiOCO accommodates both non-separable global cost functions and multi-slot feedback delay, in a heterogeneous master-worker network. 

%%%%%%%%%%%%%%%%%%%%%%%%%%%%%%%%%%%%%%%%%%%%%%%%%%%%%%%%%%%%
%           System Model & Problem Formulation            
%%%%%%%%%%%%%%%%%%%%%%%%%%%%%%%%%%%%%%%%%%%%%%%%%%%%%%%%%%%%%%

\section{System Model and Problem Formulation}
\label{Sec:Problem}

\subsection{System Model}

\begin{figure}[!t]
\centering
\vspace{0mm}
\includegraphics[width=.6\linewidth,trim=220 200 160 130,clip]{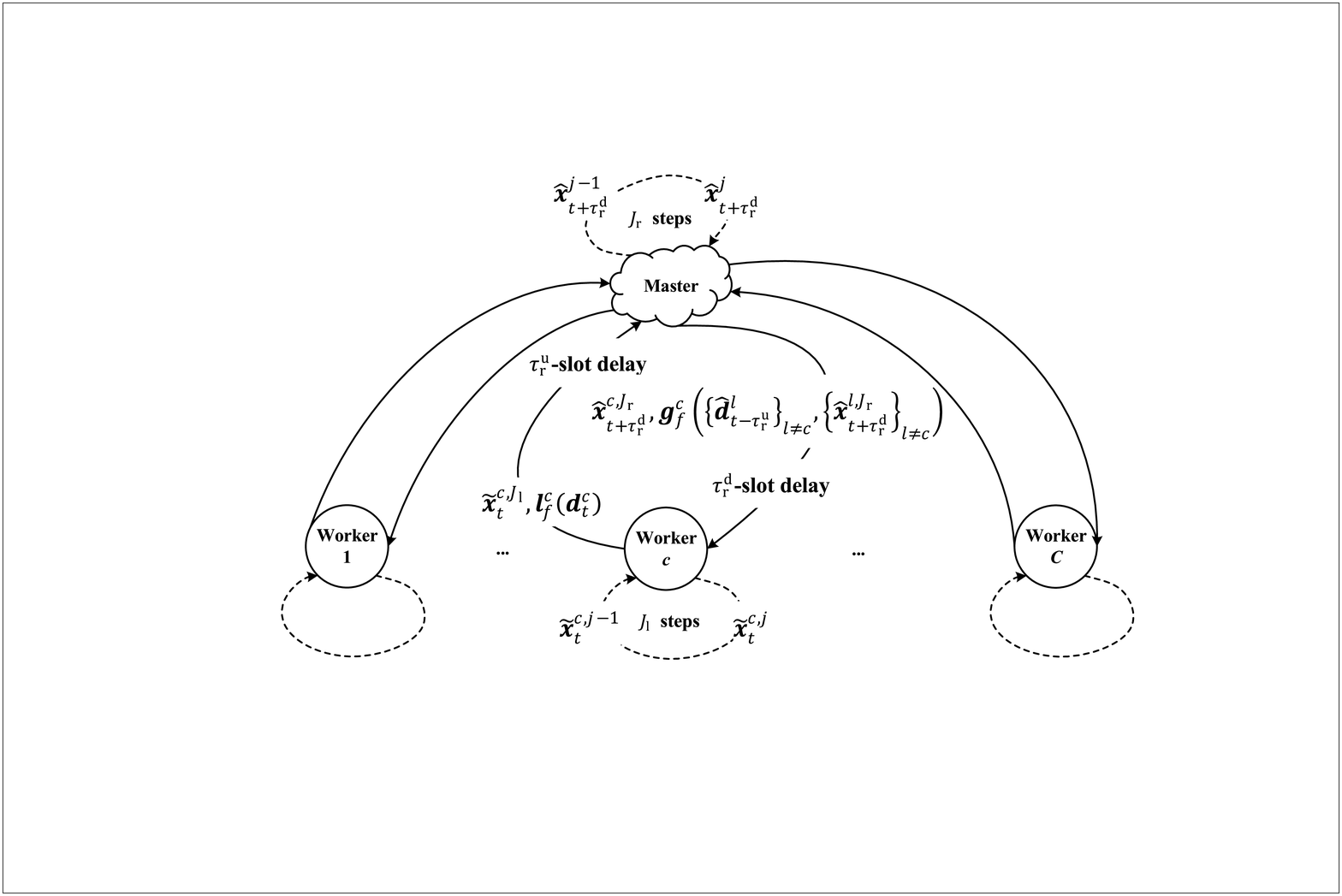}
\vspace{-6mm}
\caption {OCO over a master-worker network with communication delay (showing
the case of zero local delay for illustration).}
\label{Fig:Net}
\vspace{0mm}
\end{figure}

We consider OCO over a master-worker network in a time-slotted system with
time indexed by $t$. Since we are interested in non-separable global cost functions, it is imperative to have a master node that can collect certain global information from distributed data. As shown in  Fig.~\ref{Fig:Net}, the master node is connected to $C$ worker nodes through separate communication links. Denote by $\tau_{\text{r}}^{\text{u}}$ the \textit{remote uplink} delay for the worker nodes to upload information to the master node, and by $\tau_{\text{r}}^{\text{d}}$ the \textit{remote downlink} delay for the master node to send information back to the worker nodes.  Further denote the remote round-trip delay between the master node and the worker nodes by $\tau_{\text{r}}=\tau_{\text{r}}^{\text{u}}+\tau_{\text{r}}^{\text{d}}$.\footnote{We show later in Section \ref{Sec:HiOCO Algorithm} that only the round-trip delay impacts the online decision-making process.} In the standard OCO setting, information feedback is assumed to be delayed for only one time slot. However, in many practical applications such as those mentioned in Section \ref{sec:Introduction}, this assumption is rarely satisfied since information can be severely delayed. Therefore, in this work, we consider multi-slot delay, \ie $\tau_{\text{r}}\ge1$.
We also consider local delay denoted by $\tau_{\text{l}}$ for the worker nodes to collect their own data.  For ease of exposition, we will first consider the case $\tau_{\text{l}}=0$. Then, in Section \ref{sec:Extension with Local Delay}, we will discuss the case $\tau_{\text{l}}>0$.

At the beginning of each time slot $t$, each worker node $c$ collects some local data $\mathbf{d}_t^c$. In a large distributed network, the local data at each node often cannot be regarded as samples drawn from the same overall distribution. Furthermore, the underlying system is often time-varying in many practical applications. Therefore, we allow $\{\mathbf{d}_t^c\}_{c=1}^C$ to be non-independent and non-identically distributed and to vary arbitrarily over time with unknown statistics.

\subsection{Problem Formulation}

Let $f(\{\mathbf{d}_t^c\}_{c=1}^C,\{\mathbf{x}^c\}_{c=1}^C):\mathbb{R}^n\to\mathbb{R}$
be the \textit{global} convex cost function at time slot $t$, where $\mathbf{x}^c$ is a local decision vector at worker node~$c$, which has dimension $n^c$. We consider constraints represented by a compact convex set $\mathcal{X}^c\in\mathbb{R}^{n^c}$ at each worker node~$c$. The worker nodes and the master node cooperate to jointly select a sequence of decisions $\{\{\mathbf{x}_t^c\}_{c=1}^C\}_{t=1}^T$ from the feasible sets that minimizes the accumulated time-varying global costs. This leads to the following optimization problem:
\begin{align*}
        \textbf{P1}:~\min_{\left\{\{\mathbf{x}_t^c\in\mathcal{X}^c\}_{c=1}^C\right\}_{t=1}^T}~\sum_{t=1}^Tf\left(\{\mathbf{d}_t^c\}_{c=1}^C,\{\mathbf{x}_t^c\}_{c=1}^C\right).
\end{align*}

Existing distributed gradient descent algorithms implicitly assume each node $c$ can locally compute its own gradient $\nabla_{\mathbf{x}^c}f(\{\mathbf{d}_t^c\}_{c=1}^C,\{\mathbf{x}^c\}_{c=1}^C)$ based only on the local information \cite{Hosseini13}\nocite{Yan13}\nocite{Nedic15}\nocite{Tsianos16}\nocite{Ferdinand19}\nocite{Nunez14}\nocite{Akbari17}\nocite{Jadbabaie18}\nocite{Nima20}-\cite{Zhang19}. These algorithms focus on separable global cost functions, \ie $f(\{\mathbf{d}_t^c\}_{c=1}^C,\{\mathbf{x}^c\}_{c=1}^C)$ can be expressed as a summation of $C$ local cost functions,  each corresponding only to the local data $\mathbf{d}_t^c$ and decision vector $\mathbf{x}^c$.\footnote{ Specifically, \cite{Hosseini13}\nocite{Yan13}\nocite{Nedic15}\nocite{Tsianos16}\nocite{Ferdinand19}\nocite{Nunez14}\nocite{Akbari17}\nocite{Jadbabaie18}\nocite{Nima20}-\cite{Zhang19} assumes $f(\{\mathbf{d}_t^c\}_{c=1}^C,\mathbf{w})=\sum_{c=1}^Cf(\mathbf{d}_t^c,\mathbf{w})$, where $\mathbf{w}$ is a global decision vector. It is a special case of our global cost function $f(\{\mathbf{d}_t^c\}_{c=1}^C,\{\mathbf{x}^c\}_{c=1}^C)$ by letting $\mathbf{x}^c=\mathbf{w}$ for any $c\in\{1,\dots,C\}$.} In this work, we consider the general case where the global cost functions may be \textit{non-separable} among the worker nodes. Therefore, due to the coupling of data or variables, each worker node $c$ cannot locally compute its own gradient  without information exchange with the other nodes.

For non-separable global cost functions, the local gradient at each worker node $c$ may depend on its local data $\mathbf{d}_t^c$, local decision vector $\mathbf{x}^c$, and possibly the data $\mathbf{d}_t^l$ and decision vector $\mathbf{x}^l$ at any other worker node $l\neq{c}$. Therefore, we define the local gradient at each worker node $c$ as a general function denoted by $\mathbf{h}_f^c(\cdot)$ as follows:
\begin{align}
        \nabla_{\mathbf{x}^c}f(\{\mathbf{d}_t^c\}_{c=1}^C,\{\mathbf{x}^c\}_{c=1}^C)\triangleq\mathbf{h}_f^c\left(\mathbf{d}_t^c,\mathbf{x}^c,\mathbf{g}_f^c\left(\{\mathbf{d}_t^l\}_{l\neq{c}},\{\mathbf{x}^l\}_{l\neq{c}}\right)\right)\label{EQ:GDc}
\end{align}
where $\mathbf{g}_f^c(\cdot)$ is some global information function. Note that $\mathbf{h}_f^c(\cdot)$ and $\mathbf{g}_f^c(\cdot)$ depend on the specific format of $f(\cdot)$. Note that communicating the values of $\mathbf{g}_f^c(\cdot)$ can often reduce the communication overhead compared with sending the variables $\{\mathbf{d}_t^l\}_{l\neq{c}}$ and $\{\mathbf{x}^l\}_{l\neq{c}}$ directly.

\subsection{Performance Metric and Measure of Variation}
\label{Sec:Metric}

For notation simplicity, in the following, we define the global feasible
set as $\mathcal{X}\triangleq\cup_{c=1}^C\{\mathcal{X}^c\}\in\mathbb{R}^n$
and denote the global cost function by
\begin{align*}
        f_t(\mathbf{x})\triangleq{f}(\{\mathbf{d}_t^c\}_{c=1}^C,\{\mathbf{x}^c\}_{c=1}^C)
\end{align*}
where $\mathbf{x}\triangleq[{\mathbf{x}^1}^T,\dots,{\mathbf{\!x}^C}^T]^T\in\mathbb{R}^n$
is the global decision vector. In addition, the local gradient $\nabla_{\mathbf{x}^c}f(\{\mathbf{d}_t^c\}_{c=1}^C,\{\mathbf{x}^c\}_{c=1}^C)$
at each worker node~$c$ is denoted by $\nabla{f}_t^c(\mathbf{x}^c)$.

Due to the lack of in-time information about the global cost function at either the worker nodes or the master node, it is impossible to obtain an optimal solution to $\textbf{P1}$.\footnote{In fact, even for the most basic centralized OCO problem \cite{Zinkevich03}, an optimal solution cannot be found \cite{OT}.} Instead, we aim at selecting an online solution sequence $\{\mathbf{x}_t\}_{t=1}^T$ that keeps tracking the optimal dynamic solution sequence $\{\mathbf{x}_t^\star\}_{t=1}^T$, given by
\begin{align}
        \mathbf{x}_t^\star\in\arg\min_{\mathbf{x}\in\mathcal{X}}\{f_t(\mathbf{x})\}.\label{EQ:xt}
\end{align}
Note that $\mathbf{x}_t^\star$ is computed with the current information
about $f_t(\mathbf{x})$. The corresponding dynamic regret is defined as
\begin{align}
        \text{RE}_T^{\text{d}}\triangleq\sum_{t=1}^T\left(f_t(\mathbf{x}_t)-f_t(\mathbf{x}_t^\star)\right),\label{EQ:RE}
\end{align}
which is commonly adopted in the existing literature, \eg \cite{Jadbabaie15}\nocite{Mokhtari16}\nocite{Zhang17}\nocite{Bedi18}-\cite{Dixit19},
\cite{Jadbabaie18}\nocite{Nima20}-\cite{Zhang19}. The dynamic regret can be bounded in terms of different variation measures that represent how dynamic the system is and hence the hardness of the problem. We introduce several common variation measures as follows.

The path-length of an arbitrary sequence of reference points $\{\mathbf{r}_t\in\mathcal{X}\}_{t=1}^T$ is defined as $\Pi_T\triangleq\sum_{t=1}^T\Vert\mathbf{r}_t-\mathbf{r}_{t-1}\Vert$ \cite{Zinkevich03}. The online projected gradient descent algorithm in \cite{Zinkevich03} achieves $\mathcal{O}(\Pi_T\sqrt{T})$ dynamic regret with respect to (w.r.t.) $\{\mathbf{r}_t\}_{t=1}^T$. When the reference points are the per-slot optimal solutions, \ie $\mathbf{r}_t=\mathbf{x}_t^\star$ for all~$t$, the resulting path-length is
\begin{align}
        \Pi_T^\star\triangleq\sum_{t=1}^T\Vert\mathbf{x}_t^\star-\mathbf{x}_{t-1}^\star\Vert.\label{EQ:PiT}
\end{align}
For example, the optimistic mirror descent algorithm achieves $\mathcal{O}(\min\{\sqrt{\Gamma_{2,T}\Pi_T^\star},(\Gamma_{2,T}\Theta_TT)^{\frac{1}{3}}\})$
dynamic regret, where $\Theta_T\triangleq\sum_{t=1}^T\max_{\mathbf{x}\in\mathcal{X}}\{|f_t(\mathbf{x})-f_{t-1}(\mathbf{x})|\}$ and $\Gamma_{2,T}\triangleq\sum_{t=1}^T\Vert\nabla{f}_t(\mathbf{x}_t)-\nabla{f}_{t-1}(\mathbf{x}_{t-1})\Vert^2$ \cite{Jadbabaie15}. As another example, when the cost functions are strongly convex, the one-step gradient descent algorithm in \cite{Mokhtari16} achieves $\mathcal{O}(\Pi_T^\star)$ dynamic regret.

Another important variation measure is the squared path-length, defined as
\begin{align}
        \Pi_{2,T}^\star\triangleq\sum_{t=1}^T\Vert\mathbf{x}_t^\star-\mathbf{x}_{t-1}^\star\Vert^2.\label{EQ:Pi2T}
\end{align}
For example, the multi-step gradient descent algorithm in~\cite{{Zhang17}} improves the dynamic regret to $\mathcal{O}(\min\{\Pi_T^\star,\Pi_{2,T}^\star\})$ for strongly convex cost functions. Note that $\Pi_{2,T}^\star$ is often smaller than  $\Pi_T^\star$ in the order sense \cite{{Zhang17}}.\footnote{For instance $\Vert\mathbf{x}_t^\star-\mathbf{x}_{t-1}^\star\Vert\propto{T}^{\varrho}$ for any $t$, then $\Pi_T^\star=\mathcal{O}(T^{1+\varrho})$ and $\Pi_{2,T}^\star=\mathcal{O}(T^{1+2\varrho})$.
For a sublinear $\Pi_T^\star$ or $\Pi_{2,T}^\star$, we have $\varrho<0$ and therefore $\Pi_{2,T}^\star$ is smaller than $\Pi_T^\star$ in the order sense. Particularly, if $\varrho=-\frac{1}{2}$, we have $\Pi_{2,T}^\star=\mathcal{O}(1)$ and $\Pi_T^\star=T^{\frac{1}{2}}$.}

Further variation measures are required when we use inexact gradients. For example, the standard and proximal online gradient descent algorithms were respectively extended in \cite{Bedi18} and \cite{Dixit19} to accommodate inexact gradients. Both achieve $\mathcal{O}\left(\max\{\Pi_T^\star,\Delta_T\}\right)$ dynamic regret, where $\Delta_T$ is the accumulated gradient error defined as
\begin{align}
        \Delta_T\triangleq\sum_{t=1}^T\max_{\mathbf{x}\in\mathcal{X}}\Vert\nabla{f}_t(\mathbf{x})-\nabla\hat{f}_t(\mathbf{x})\Vert\label{EQ:DeltaT}
\end{align}
where $\nabla\hat{f}_t(\cdot)$ is a given function available at the decision
maker to predict the current gradient.

For the performance bounding of HiOCO in Section \ref{Sec:Regret}, we further define the accumulated squared
gradient error as 
\begin{align}
        \Delta_{2,T}\triangleq\sum_{t=1}^T\max_{\mathbf{x}\in\mathcal{X}}\{\Vert\nabla{f}_t(\mathbf{x})-\nabla\hat{f}_t(\mathbf{x})\Vert^2\}.\label{EQ:Delta2T}
\end{align}
Similar to the relationship between $\Pi_{2,T}^\star$ and  $\Pi_T^\star$, $\Delta_{2,T}$ is often smaller than $\Delta_T$ in the order sense.

The above works \cite{Zinkevich03}, \cite{Jadbabaie15}\nocite{Mokhtari16}\nocite{{Zhang17}}\nocite{Bedi18}-\cite{Dixit19} focus on centralized OCO. For distributed OCO, the mirror descent algorithm in \cite{Jadbabaie18} achieves $\mathcal{O}(\sqrt{\Pi_T^\star{T}})$ dynamic regret. The any-bach mirror descent algorithm in \cite{Nima20} accommodates time-varying batch sizes while recovering the dynamic regret in \cite{Jadbabaie18} as a special case.  The impact of strongly convex cost functions on the dynamic regret of distributed OCO has been considered in \cite{Zhang19}. The push-pull based online algorithm provides $\mathcal{O}(\Pi_T^\star,\Gamma_T)$ dynamic regret, where $\Gamma_T\triangleq\sum_{t=1}^T\Vert\nabla{f}_{t}(\mathbf{x}_t)-\nabla{f}_{t-1}(\mathbf{x}_{t-1})\Vert_\infty$. In this work, we will derive dynamic regret bounds using variation measures defined in (\ref{EQ:PiT})-(\ref{EQ:Delta2T}).

\section{Hierarchical Online Convex Optimization}
\label{Sec:HiOCO}

In this section, we first present the design details of HiOCO. We then study the impact of hierarchical multi-step estimated gradient descent on the performance of HiOCO in terms of its dynamic regret. We further give sufficient conditions under which HiOCO yields sublinear dynamic regret. Finally, we discuss the performance merits of HiOCO over existing OCO algorithms in terms of the dynamic regret bound.

\subsection{HiOCO Algorithm  Description}
\label{Sec:HiOCO Algorithm}

Existing distributed OCO algorithms \cite{Hosseini13}\nocite{Yan13}\nocite{Nedic15}\nocite{Tsianos16}\nocite{Ferdinand19}\nocite{Nunez14}\nocite{Akbari17}\nocite{Jadbabaie18}\nocite{Nima20}-\cite{Zhang19} cannot be directly applied to solve \textbf{P1} with non-separable global cost functions. As an alternative, one may apply centralized OCO \cite{Zinkevich03}\nocite{OT}\nocite{Slow}\nocite{AD}\nocite{Hall15}\nocite{Jadbabaie15}\nocite{Mokhtari16}\nocite{Zhang17}\nocite{Bedi18}-\cite{Dixit19} at the master node after it has received all the local data from the worker nodes. However, this way of solving the problem does not take advantage of the more timely local information at the worker nodes, nor their computation resources.

Different from existing centralized and distributed OCO algorithms, in HiOCO, the master node and the worker nodes cooperate in gradient estimation and decision updates, by taking full advantage of the network heterogeneity in information timeliness and computation capacity. For ease of exposition, we first consider the case of zero local delay at the worker nodes but will extend that to the case of non-zero local delay in Section~\ref{sec:Extension with Local Delay}. In the following, we present HiOCO algorithms at the master node and the worker nodes.

\subsubsection{Master Node's Algorithm}

In practical systems, the master node often has a higher computation capacity compared with the worker nodes. To leverage this capacity, we design HiOCO to be capable of performing central gradient descent at the master node. In this case, each worker node $c$ needs to share information about its local data $\mathbf{d}_t^c$ with the master node. Each worker node $c$ sends a compressed version of the current local data $\mathbf{l}_f^c(\mathbf{d}_t^c)$ to the master node, where $\mathbf{l}_f^c(\cdot)$ is some general function for data compression. Note that the need for information exchange
about the local data is unavoidable in order to optimize non-separable global
cost functions, regardless whether a master node is used.

At the beginning of each time slot $t$, each worker node $c$ executes its current local decision vector $\mathbf{x}_t^c$ and then uploads it together with $\mathbf{l}_f^c(\mathbf{d}_t^c)$ to the master node. Due to the remote uplink delay, at the beginning of each time slot $t$, the master node only has the $\tau_{\text{r}}^{\text{u}}$-slot-delayed local decision vector $\mathbf{x}_{t-\tau_{\text{r}}^{\text{u}}}^c$ and compressed data $\mathbf{l}_f^c(\mathbf{d}_{t-\tau_{\text{r}}^{\text{u}}}^c)$ from each worker node~$c$. The master node then recovers an estimated version of the local data $\hat{\mathbf{d}}_{t-\tau_{\text{r}}^{\text{u}}}^c$ from $\mathbf{l}_f^c(\mathbf{d}_{t-\tau_{\text{r}}^{\text{u}}}^c)$, which is then used to generate new decision vectors to assist the local decision-making processes at the worker nodes.\footnote{The compression and recovery techniques on the data can be chosen based on specific applications and are beyond the scope of this paper.}

Note that the master node needs to consider the remote downlink delay and design the decision vectors for the worker nodes $\tau_{\text{r}}^{\text{d}}$-slot ahead based on the $\tau_{\text{r}}^{\text{u}}$-slot delayed information. One can easily verify that only the round-trip remote delay $\tau_{\text{r}}$ impacts the decision-making process. Therefore, in the following, without loss of generality, we only need to consider the case of $\tau_{\text{r}}$-slot remote uplink delay and zero remote downlink delay.

\begin{remark}
There is often a delay-accuracy tradeoff for the recovered data $\{\hat{\mathbf{d}}_{t-\tau_{\text{r}}}^c\}_{c=1}^C$ at the master node, since a higher data compression rate reduces the data transmission time but also reduces the data accuracy at the master node. If data privacy is a concern, the worker nodes can also add noise to the compressed data while sacrificing some system performance~\cite{AbediDP16}.
\end{remark}

With $\{\mathbf{x}_{t-\tau_{\text{r}}}^c\}_{c=1}^C$ and $\{\hat{\mathbf{d}}_{t-\tau_{\text{r}}}^c\}_{c=1}^C$, for each worker node $c$, the master node sets an intermediate decision vector $\hat{\mathbf{x}}_{t}^{c,0}=\mathbf{x}_{t-\tau_{\text{r}}}^c$ and performs $J_{\text{r}}$-step gradient descent to generate $\hat{\mathbf{x}}_{t}^{c,J_{\text{r}}}$.\footnote{Later in Sections \ref{Sec:Regret},  we show that multi-step gradient descent in HiOCO improves the dynamic regret bound.} For each gradient descent step $j\in\{1,\dots,J_{\text{r}}\}$, the master node solves the following optimization problem for $\hat{\mathbf{x}}_{t}^{c,j}$:
\begin{align*}
        \textbf{P2}:\min_{\mathbf{x}^c\in\mathcal{X}^c}\langle\nabla\hat{f}_{t-\tau_{\text{r}}}^c(\hat{\mathbf{x}}_{t}^{c,j-1}),\mathbf{x}^c-\hat{\mathbf{x}}_{t}^{c,j-1}\rangle+\frac{\alpha}{2}\Vert\mathbf{x}^c-\hat{\mathbf{x}}_{t}^{c,j-1}\Vert^2
\end{align*}
where $\nabla\hat{f}_{t-\tau_{\text{r}}}^c(\hat{\mathbf{x}}_{t}^{c,j-1})$ is an \textit{estimated} gradient based on the delayed global information $\{\hat{\mathbf{d}}_{t-\tau_{\text{r}}}^c\}_{c=1}^C$, and it is given by
\begin{align}
        \nabla\hat{f}_{t-\tau_{\text{r}}}^c(\hat{\mathbf{x}}_{t}^{c,j-1})\triangleq\!\mathbf{h}_f^c\!\left(\hat{\mathbf{d}}_{t-\tau_{\text{r}}}^c,\hat{\mathbf{x}}_{t}^{c,j-1}\!,\mathbf{g}_f^c\!\left(\!\{\hat{\mathbf{d}}_{t-\tau_{\text{r}}}^l\}_{l\neq{c}},\{\hat{\mathbf{x}}_{t}^{l,j-1}\}_{l\neq{c}}\right)\right)\!.\!\!\!\!\label{EQ:EGDmaster}\!\!
\end{align}

The master node then sends $\hat{\mathbf{x}}_{t}^{c,J_{\text{r}}}$ and the  global information $\mathbf{g}_f^c(\{\hat{\mathbf{d}}_{t-\tau_{\text{r}}}^l\}_{l\neq{c}},\{\hat{\mathbf{x}}_{t}^{l,J_{\text{r}}}\}_{l\neq{c}})$ to assist the local decision-making process at each worker node $c$. In Algorithm~\ref{Alg:Master}, we summarize the master node's algorithm.

\begin{algorithm}[!t]
\caption{HiOCO master node's algorithm}
\label{Alg:Master}
\begin{algorithmic}[1]
\STATE {Initialize $\alpha>0$ and broadcast it to each worker node $c$. }
\STATE{At the beginning of each $t>\tau_{\text{r}}$, do the following:}
\STATE{Receive $\mathbf{x}_{t-\tau_{\text{r}}}^c$ and $\mathbf{l}_f^c(\mathbf{d}_{t-\tau_{\text{r}}}^c)$
from each worker node~$c$.}
\STATE{Estimate $\hat{\mathbf{d}}_{t-\tau_{\text{r}}}^c$ from $\mathbf{l}_f^c(\mathbf{d}_{t-\tau_{\text{r}}}^c)$.}
\STATE{Set $\hat{\mathbf{x}}_{t}^{c,0}=\mathbf{x}_{t-\tau_{\text{r}}}^c$
for each worker node $c$.\\}
\STATE{\textbf{for} $j=1$ \textbf{to} $J_{\text{r}}$}\\
\STATE \quad Construct estimated gradient $\nabla\hat{f}_{t-\tau_{\text{r}}}^c(\hat{\mathbf{x}}_{t}^{c,j-1})$
in (\ref{EQ:EGDmaster}).
\STATE{\quad Update $\hat{\mathbf{x}}_{t}^{c,j}$ for each  worker node
$c$ by solving \textbf{P2}.}
\STATE{\textbf{end for}}
\STATE{Send $\hat{\mathbf{x}}_{t}^{c,J_{\text{r}}}$ and  $\mathbf{g}_f^c(\{\hat{\mathbf{d}}_{t-\tau_{\text{r}}}^l\}_{l\neq{c}},\{\hat{\mathbf{x}}_{t}^{l,J_{\text{r}}}\}_{l\neq{c}})$
to each worker node $c$.}
\end{algorithmic}
\end{algorithm}

\begin{remark}
Note that even though the master node has global information about the data, this information is delayed and inexact. In HiOCO, different from the centralized approaches \cite{Zinkevich03}\nocite{OT}\nocite{Slow}\nocite{AD}\nocite{Hall15}\nocite{Jadbabaie15}\nocite{Mokhtari16}\nocite{Zhang17}\nocite{Bedi18}-\cite{Dixit19}, the central decisions made at the master node are not used directly as the final solution, but are used later at the worker nodes to assist their local decision-making processes.
\end{remark}

\subsubsection{Worker Node $c$'s Algorithm} 

The worker nodes have the most up-to-date information about their own local data. However, when the global cost function is non-separable, each worker node $c$ cannot compute  its own gradient $\nabla{f}_t^c(\mathbf{x}_{t}^c)=\mathbf{h}_f^c(\mathbf{d}_t^c,\mathbf{x}_t^c,\mathbf{g}_f^c(\{\mathbf{d}_t^l\}_{l\neq{c}},\{\mathbf{x}_t^l\}_{l\neq{c}}))$ based only on the local information. Therefore, in HiOCO, the master node assists the local gradient estimation by communicating the required global information $\mathbf{g}_f^c(\{\hat{\mathbf{d}}_{t-\tau_{\text{r}}}^l\}_{l\neq{c}},\{\hat{\mathbf{x}}_{t}^{l,J_{\text{r}}}\}_{l\neq{c}})$ to each worker node $c$. Note that due to the communication delay and data compression, such global information received by each worker node $c$ is delayed and inexact. 

At the beginning of each time slot $t$, each worker node~$c$ receives the global information $\mathbf{g}_f^c(\{\hat{\mathbf{d}}_{t-\tau_{\text{r}}}^l\}_{l\neq{c}},\{\hat{\mathbf{x}}_{t}^{l,J_{\text{r}}}\}_{l\neq{c}})$
together with the intermediate decision vector $\hat{\mathbf{x}}_{t}^{c,J_{\text{r}}}$
from the master node. Each worker node $c$ then initializes another intermediate decision vector $\tilde{\mathbf{x}}_t^{c,0}=\hat{\mathbf{x}}_t^{c,J_{\text{r}}}$
and performs $J_{\text{l}}$-step local gradient descent to generate $\tilde{\mathbf{x}}_t^{c,J_{\text{l}}}$. For each gradient descent step $j\in\{1,\dots,J_{\text{l}}\}$, each worker node $c$ solves the following optimization problem for $\tilde{\mathbf{x}}_t^{c,j}$:
\begin{align*}
        \textbf{P3}:\min_{\mathbf{x}^c\in\mathcal{X}^c}\langle\nabla\hat{f}_{t}^c(\tilde{\mathbf{x}}_{t}^{c,j-1}),\mathbf{x}^c-\tilde{\mathbf{x}}_{t}^{c,j-1}\rangle+\frac{\alpha}{2}\Vert\mathbf{x}^c-\tilde{\mathbf{x}}_{t}^{c,j-1}\Vert^2
\end{align*}
where $\nabla\hat{f}_{t}^c(\tilde{\mathbf{x}}_{t}^{c,j-1})$ is an \textit{estimated} gradient based on the timely local data $\mathbf{d}_t^c$ and the delayed global information $\mathbf{g}_f^c(\{\hat{\mathbf{d}}_{t-\tau_{\text{r}}}^l\}_{l\neq{c}},\{\hat{\mathbf{x}}_{t}^{l,J_{\text{r}}}\}_{l\neq{c}})$, and it is given by
\begin{align}
        \nabla\hat{f}_{t}^c(\tilde{\mathbf{x}}_{t}^{c,j-1})\triangleq\mathbf{h}_f^c\left(\mathbf{d}_t^c,\tilde{\mathbf{x}}_t^{c,j-1},\mathbf{g}_f^c\left(\{\hat{\mathbf{d}}_{t-\tau_{\text{r}}}^l\}_{l\neq{c}},\{\hat{\mathbf{x}}_t^{l,J_{\text{r}}}\}_{l\neq{c}}\right)\right)\!.\!\label{EQ:EGDworkerc}\!\!
\end{align}

The above estimated gradient takes full advantage of the information timeliness at the worker nodes, as well as the central availability of information at the master node, to enable local gradient descent at the worker nodes for non-separable global cost minimization. Each worker node $c$ executes $\mathbf{x}_t^c=\tilde{\mathbf{x}}_t^{c,J_{\text{l}}}$ as its current local decision vector. It then uploads $\mathbf{x}_t^c$ and the compressed local data $\mathbf{l}_f^c(\mathbf{d}_t^c)$ to the master node. In Algorithm \ref{Alg:Worker}, we summarize the worker node's algorithm.

\begin{algorithm}[!t]
\caption{HiOCO worker node $c$'s algorithm}
\label{Alg:Worker}
\begin{algorithmic}[1]
\STATE Initialize $\mathbf{x}_t^c\in\mathcal{X}^c$ at random for any ${t}\le\tau_{\text{r}}$.
\STATE At the beginning of each $t>\tau_{\text{r}}$, do the following:
\STATE Receive $\hat{\mathbf{x}}_t^{c,J_{\text{r}}}$ and $\mathbf{g}_f^c(\{\hat{\mathbf{d}}_{t-\tau_{\text{r}}}^l\}_{l\neq{c}},\{\hat{\mathbf{x}}_t^{l,J_{\text{r}}}\}_{l\neq{c}})$
from the master node.
\STATE Set $\tilde{\mathbf{x}}_t^{c,0}=\hat{\mathbf{x}}_t^{c,J_{\text{r}}}$.
\STATE \textbf{for} $j=1$ \textbf{to} $J_{\text{l}}$
\STATE \quad Construct estimated gradient $\nabla\hat{f}_{t}^c(\tilde{\mathbf{x}}_{t}^{c,j-1})$
in (\ref{EQ:EGDworkerc}).
\STATE \quad Update $\tilde{\mathbf{x}}_t^{c,j}$ by solving \textbf{P3}.
\STATE{\textbf{end for}}
\STATE Set $\mathbf{x}_t^c=\tilde{\mathbf{x}}_t^{c,J_{\text{l}}}$ and execute
$\mathbf{x}_t^c$.
\STATE Send $\mathbf{x}_t^c$ and $\mathbf{l}_f^c(\mathbf{d}_{t}^c)$ to the
master node.
\end{algorithmic}
\end{algorithm}

\begin{remark}
The solutions to \textbf{P2} and \textbf{P3} are projected gradient descent updates. For example, solving \textbf{P2} for $\hat{\mathbf{x}}_{t}^{c,j}$ is equivalent to updating $\hat{\mathbf{x}}_{t}^{c,j}$ through
\begin{align*}
        \hat{\mathbf{x}}_{t}^{c,j}=\mathcal{P}_{\mathcal{X}^c}\left\{\hat{\mathbf{x}}_{t}^{c,j-1}-\frac{1}{\alpha}\nabla\hat{f}_{t-\tau_{\text{r}}}^c(\hat{\mathbf{x}}_{t}^{c,j-1})\right\}
\end{align*}
where $\mathcal{P}_{\mathcal{X}^c}\{\mathbf{x}^c\}\triangleq\arg\min_{\mathbf{y}^c\in\mathcal{X}^c}\{\Vert\mathbf{y}^c-\mathbf{x}^c\Vert^2\}$
is the projection operator onto the convex feasible set $\mathcal{X}^c$ and
$\alpha$ can be seen as a step-size parameter. We use \textbf{P2} and \textbf{P3} here for the ease of dynamic regret analysis later.
\end{remark}

\begin{remark}
For separable global cost functions, HiOCO can still be applied. In this case, it is still beneficial to perform central gradient descent for improved system performance, while incurring some communication overhead caused by uploading the compressed local data. \end{remark}

\begin{remark}
Single-step and multi-step gradient descent algorithms were provided in \cite{Mokhtari16} and \cite{Zhang17}, while \cite{Bedi18} and \cite{Dixit19} proposed single-step inexact gradient descent algorithms. All of these algorithms are centralized and are under the standard OCO setting with one-slot delay. In HiOCO, both the master node and the worker nodes can perform multi-step gradient descent with estimated gradients under multi-slot delay.
\end{remark}

\subsection{Dynamic Regret Bounds}
\label{Sec:Regret}

In this section, we derive upper bounds on the dynamic regret of HiOCO. We develop new analysis techniques to account for its hierarchical multi-step gradient descent with estimated gradients, in the presence of multi-slot delay.

We focus on strongly convex functions, which arise in many machine learning and signal processing applications, such as Lasso regression, support vector machine, softmax classifier, and robust subspace tracking. Furthermore, for applications with general convex cost functions, it is common to add a simple regularization term such as $\frac{\mu}{2}\Vert\mathbf{x}\Vert^2$, so that the overall optimization objective becomes strongly convex \cite{Dixit19}. We make the following assumptions that are common in the literature of OCO with strongly convex functions \cite{Mokhtari16}\nocite{Zhang17}\nocite{Bedi18}-\cite{Dixit19}, \cite{Zhang19}.

\begin{assumption}\label{Asp:f}
For any $t$, $f_t(\mathbf{x})$ satisfies the following:
\begin{enumerate}
\item[1.1)]  $f_t(\mathbf{x})$ is $\mu$-strongly convex over $\mathcal{X}$, \ie $\exists\mu>0$, \st for any ${\mathbf{x},\mathbf{y}}\in\mathcal{X}$ and $t$
\begin{align}
        f_t(\mathbf{y})\ge{f}_t(\mathbf{x})+\langle\nabla{f}_t(\mathbf{x}),\mathbf{y}-\mathbf{x}\rangle+\frac{\mu}{2}\Vert\mathbf{y}-\mathbf{x}\Vert^2.\label{EQ:SConvex}
\end{align}
\item[1.2)] $f_t(\mathbf{x})$ is $L$-smooth over $\mathcal{X}$, \ie $\exists{L}>0$, \st for any $\mathbf{x},\mathbf{y}\in\mathcal{X}$ and $t$
\begin{align}
        f_t(\mathbf{y})\le{f}_t(\mathbf{x})+\langle\nabla{f}_t(\mathbf{x}),\mathbf{y}-\mathbf{x}\rangle+\frac{L}{2}\Vert\mathbf{y}-\mathbf{x}\Vert^2.\label{EQ:Smooth}
\end{align}

\item[1.3)] The gradient of $f_t(\mathbf{x})$ is bounded, \ie $\exists{D}>0$,
\st for any $\mathbf{x}\in\mathcal{X}$ and $t$
\begin{align}
        \Vert\nabla{f}_t(\mathbf{x})\Vert\le{D}.\label{EQ:D}
\end{align}
\end{enumerate}
\end{assumption}

\begin{assumption}\label{Asp:x}
The radius of $\mathcal{X}$ is bounded, \ie $\exists{R}\!>\!0$, \st for any $\mathbf{x},\mathbf{y}\in\mathcal{X}$
\begin{align}
        \Vert\mathbf{x}-\mathbf{y}\Vert\le{R}.\label{EQ:R}
\end{align}
\end{assumption}

We require the following lemma, which is reproduced from Lemma 2.8 in \cite{OLBK}. 

\begin{lemma}\label{lm:StrConv}
Let $\mathcal{X}\in\mathbb{R}^n$ be a nonempty convex set. Let $f(\mathbf{x}):\mathbb{R}^n\to\mathbb{R}$
be a $\mu$-strongly-convex function over $\mathcal{X}$. Let $\mathbf{x}^\star\in\arg\min_{\mathbf{x}\in\mathcal{X}}\{f(\mathbf{x})\}$. Then, for any $\mathbf{y}\in\mathcal{X}$, we have
\begin{align*}
        f(\mathbf{x}^\star)\le{f}(\mathbf{y})-\frac{\mu}{2}\Vert\mathbf{y}-\mathbf{x}^\star\Vert^2.
\end{align*}
\end{lemma}

To proceed with our analysis, we first need to quantify the impact of one-step estimated gradient descent in terms of the squared gradient estimation error. This is given in the following lemma.

\begin{lemma}\label{lm:GDE}
Assume that $f(\mathbf{x}):\mathcal{X}\to\mathbb{R}$ is a $\mu$-strongly convex and $L$-smooth function. Let $\mathbf{z}\in\arg\min_{\mathbf{x}\in\mathcal{X}}\{\langle\nabla\hat{f}(\mathbf{y}),\mathbf{x}-\mathbf{y}\rangle+\frac{\alpha}{2}\Vert\mathbf{x}-\mathbf{y}\Vert^2\}$,
where $\nabla\hat{f}(\mathbf{y})$ is an estimated gradient of $\nabla{f}(\mathbf{y})$,
and $\mathbf{x}^\star\in\arg\min_{\mathbf{x}\in\mathcal{X}}\{f(\mathbf{x})\}$. For any $\alpha\ge{L}$, and $\gamma\in(0,2\mu)$, we have
\begin{align}
        \Vert\mathbf{z}-\mathbf{x}^\star\Vert^2\le\eta\Vert\mathbf{y}-\mathbf{x}^\star\Vert^2+\beta\Vert\nabla\hat{f}(\mathbf{y})-\nabla{f}(\mathbf{y})\Vert^2\label{eq:GDE-1}
\end{align}
where $\eta=\frac{\alpha-\mu}{\alpha+\mu-\gamma}<1$ and $\beta=\frac{1}{\gamma(\alpha+\mu-\gamma)}$.
\end{lemma}
\textit{Proof:} See Appendix \ref{app:GDE}.

\begin{remark}
From (\ref{eq:GDE-1}), the sufficient condition for $\Vert\mathbf{z}-\mathbf{x}^\star\Vert^2\le\Vert\mathbf{y}-\mathbf{x}^\star\Vert^2$ is $\Vert\nabla\hat{f}(\mathbf{y})-\nabla{f}(\mathbf{y})\Vert^2<\gamma(2\mu-\gamma)\Vert\mathbf{y}-\mathbf{x}^\star\Vert^2$. This condition on the gradient estimation error is most easily satisfied when $\gamma=\mu$. In this case, the contraction constant $\eta=\frac{\alpha-\mu}{\alpha}$ recovers the one in \cite{Mokhtari16}. Furthermore, as $\gamma$ approaches $0$, $\eta$ approaches the contraction constant $\frac{\alpha-\mu}{\alpha+\mu}$ in \cite{Zhang17}. Different from Proposition 2 in \cite{Mokhtari16} and Lemma 5 in \cite{Zhang17}, Lemma \ref{lm:GDE} takes into account the impacts of estimated gradient descent and recovers the results in \cite{Mokhtari16} and \cite{Zhang17} as special cases.
\end{remark}

\begin{remark}

To show a contraction relationship, the optimal gradient descent step-size in \cite{Bedi18} needs to be in a specific range based on the knowledge of $\mu$ in (\ref{EQ:SConvex}), $L$ in (\ref{EQ:Smooth}), and $\nu$ from an additional assumption $\Vert\nabla\hat{f}_t(\mathbf{x}_t)-\nabla{f}_t(\mathbf{x}_t)\Vert^2\le\epsilon^2+\nu^2\Vert\nabla{f}_t(\mathbf{x}_t)\Vert^2$ for some $\epsilon\ge0$ and $\nu\ge0$.  The contraction analysis in \cite{Dixit19} focuses on the proximal point algorithm and is substantially different from Lemma~\ref{lm:GDE}.
\end{remark}

Leveraging Lemmas \ref{lm:StrConv} and \ref{lm:GDE},  we examine the impact of hierarchical multi-step gradient descent with estimated gradients  on the dynamic regret bounds for OCO. The following theorem provides two upper bounds on the dynamic regret $\text{RE}_T^{\text{d}}$ for HiOCO.

\begin{theorem} \label{thm}
For any $\alpha\ge{L}$, $\xi>0$ and $\gamma\in(0,2\mu)$, the dynamic regret yielded by HiOCO is bounded as follows:

\begin{enumerate}

\item[\romannum{1})] For any  $J_{\text{l}}+J_{\text{r}}\ge1$, we have
\begin{align*}
        \text{RE}_T^{\text{d}}\le\tau_{\text{r}}{DR}+\frac{D}{1-\sqrt{\eta^{J_{\text{l}}+J_{\text{r}}}}}\left(\tau_{\text{r}}{R}+\tau_{\text{r}}\Pi_T^\star+\frac{\sqrt{\beta}}{1-\sqrt{\eta}}\Delta_T\right)\!.
\end{align*}

\item[\romannum{2})] For any $J_{\text{l}}+J_{\text{r}}\ge1$ such that $2\eta^{J_{\text{l}}+J_{\text{r}}}<1$,
we have
\begin{align*}
        \text{RE}_T^{\text{d}}&\le\frac{1}{2\xi}\sum_{t=1}^T\Vert\nabla{f}_t(\mathbf{x}_t^\star)\Vert^2+\frac{L+\xi}{2}\tau_{\text{r}}{R}^2+\frac{L+\xi}{2(1-2\eta^{J_{\text{l}}+J_{\text{r}}})}\left(\tau_{\text{r}}{R}^2+2\tau_{\text{r}}^2\Pi_{2,T}^\star+\frac{2\beta}{1-\eta}\Delta_{2,T}\right)\!.
\end{align*}

\end{enumerate}
\end{theorem}
\textit{Proof:} See Appendix \ref{app:thm}.

\begin{remark}
The dynamic regret bounds in Theorem \ref{thm} hold for any general gradient estimation schemes that can be used in HiOCO. Note that in the definitions of $\Delta_T$ in (\ref{EQ:DeltaT}) and $\Delta_{2,T}$ in (\ref{EQ:Delta2T}),  $\max_{\mathbf{x}\in\mathcal{X}}\{\Vert\nabla{f}_t(\mathbf{x})-\nabla\hat{f}_t(\mathbf{x})\Vert\}$
is the maximum gradient estimation error w.r.t. some general gradient estimation function $\nabla\hat{f}(\cdot)$. Therefore, it serves as an upper bound for the gradient estimations in (\ref{EQ:EGDmaster}) and (\ref{EQ:EGDworkerc}). \end{remark}

\subsection{Discussion on the Dynamic Regret Bounds}

In this section, we discuss the sufficient conditions for HiOCO to yield
sublinear dynamic regret and highlight several prominent advantages of HiOCO over existing OCO algorithms. 

We note that $\sum_{t=1}^T\Vert\nabla{f}_t(\mathbf{x}_t^\star)\Vert^2$ is often small. In particular, if $\mathbf{x}_t^\star$ is an interior point of $\mathcal{X}$ or \textbf{P1} is an unconstrained online problem, we readily have $\nabla{f}_t(\mathbf{x}_t^\star)=\mathbf{0}$.  Therefore, we usually have $\sum_{t=1}^T\Vert\nabla{f}_t(\mathbf{x}_t^\star)\Vert^2=\mathcal{O}(\min\{\Pi_T^\star,\Pi_{2,T}^\star\})$~\cite{Zhang17}. In this case, we can simplify Theorem~\ref{thm} to the following:

\begin{corollary}\label{Cor-BD}
Suppose $\!\sum_{t=1}^T\!\!\Vert\nabla{f}_t(\mathbf{x}_t^\star)\Vert^2\!\!=\!\mathcal{O}(\min\{\Pi_{T}^\star,\!\Pi_{2,T}^\star\}\!)$. For any $J_{\text{l}}+J_{\text{r}}\ge1$ such that $2\eta^{J_{\text{l}}+J_{\text{r}}}<1$, we have $\text{RE}_T^{\text{d}}=\mathcal{O}(\min\{\max\{\tau_{\text{r}}\Pi_T^\star,\Delta_{T}\},\max\{\tau_{\text{r}}^2\Pi_{2,T}^\star,\Delta_{2,T}\}\})$.
\end{corollary}

Note that the feedback delay is always bounded by some constant in practice, \ie $\tau_{\text{r}}=\mathcal{O}(1)$. Thus, from Corollary \ref{Cor-BD}, a sufficient condition for HiOCO to yield sublinear dynamic regret is either $\max\{\Pi_T^\star,\Delta_{T}\}=\mathbf{o}(T)$ or $\max\{\Pi_{2,T}^\star,\Delta_{2,T}\}=\mathbf{o}(T)$, \ie the  variation measures grow sublinearly over time. 

\begin{remark}
Sublinearity of the  variation measures is necessary to have sublinear dynamic regret \cite{Gap}. This can be seen from the dynamic regret bounds derived in \cite{Zinkevich03}, \cite{Jadbabaie15}\nocite{Mokhtari16}\nocite{Zhang17}\nocite{Bedi18}-\cite{Dixit19},
\cite{Jadbabaie18}\nocite{Nima20}-\cite{Zhang19} as shown in Table \ref{Tab:RE}. In many online applications,
the system tends to stabilize over time, leading to sublinear system variation and thus sublinear dynamic regret.
\end{remark}

We show in the following that the dynamic regret of HiOCO recovers or improves over the ones in \cite{Mokhtari16}\nocite{Zhang17}\nocite{Bedi18}-\cite{Dixit19}.\footnote{Existing works on distributed OCO \cite{Hosseini13}\nocite{Yan13}\nocite{Nedic15}\nocite{Tsianos16}\nocite{Ferdinand19}\nocite{Nunez14}\nocite{Akbari17}\nocite{Jadbabaie18}\nocite{Nima20}-\cite{Zhang19} are limited to separable cost functions and thus cannot solve our online problem. Therefore, the dynamic regret bounds derived in \cite{Hosseini13}\nocite{Yan13}\nocite{Nedic15}\nocite{Tsianos16}\nocite{Ferdinand19}\nocite{Nunez14}\nocite{Akbari17}\nocite{Jadbabaie18}\nocite{Nima20}-\cite{Zhang19}
are not comparable with ours.} 

\begin{remark}
The centralized single-step and multi-step gradient descent algorithms achieve
$\mathcal{O}(\Pi_T^\star)$ and $\mathcal{O}(\min\{\Pi_T^\star,\Pi_{2,T}^\star\})$ dynamic regret in \cite{Mokhtari16} and \cite{Zhang17}, respectively. If we configure HiOCO to perform gradient descent only at the master node, and assume one-slot delayed accurate data as in \cite{Mokhtari16} and \cite{Zhang17}, the resulting dynamic regrets recover the ones in \cite{Mokhtari16} and \cite{Zhang17} as special cases.
\end{remark}

\begin{remark}
The centralized single-step inexact gradient descent algorithms in \cite{Bedi18} and \cite{Dixit19} achieves $\mathcal{O}\left(\max\{\Pi_T^\star,\Delta_T\}\right)$ dynamic regret with one-slot delay. By configuring HiOCO to perform single-step gradient descent only at the master node, and assuming one-slot delayed inexact data as in \cite{Bedi18} and \cite{Dixit19}, HiOCO  recovers the dynamic regret bound in \cite{Bedi18} and \cite{Dixit19} as a special case. If the master node performs multi-step gradient descent, HiOCO yields an improved $\mathcal{O}(\min\{\max\{\Pi_T^\star,\Delta_{T}\},\max\{\Pi_{2,T}^\star,\Delta_{2,T}\}\})$ dynamic regret compared with  \cite{Bedi18} and \cite{Dixit19}.
\end{remark}

\section{Extension to Non-Zero Local Delay}
\label{sec:Extension with Local Delay}

We now consider the case of non-zero local delay, \ie at the beginning of
each time slot $t$, each worker node $c$ only has the $\tau_{\text{l}}$-delayed
local data $\mathbf{d}_{t-\tau_{\text{l}}}^c$ for some $\tau_{\text{l}}>0$. Let $\tau=\tau_{\text{l}}+\tau_{\text{r}}$ be the total delay. We extend Algorithms~\ref{Alg:Master} and \ref{Alg:Worker} to deal with non-zero local delay as follows.

In Algorithm \ref{Alg:Master}, we make the following modifications: \romannum{1}) start the algorithm at $t>\tau$ in Step 2; \romannum{2}) change $\mathbf{d}_{t-\tau_{\text{r}}}^c$ to $\mathbf{d}_{t-\tau}^c$ in Steps 2 and 3; \romannum{3}) modify $\hat{\mathbf{d}}_{t-\tau_{\text{r}}}^c$ to $\hat{\mathbf{d}}_{t-\tau}^c$ in Steps 4 and 10; \romannum{4}) set $\hat{\mathbf{x}}_t^{c,0}=\mathbf{x}_{t-\tau}^c$ in Step 5;\footnote{A more recent decision vector $\mathbf{x}_{t-\tau_{\text{l}}}^{c}$
than the dataset $\hat{\mathbf{d}}_{t-\tau}^c$ at the master node does not
help to make a more accurate gradient estimation. Therefore, the  timeliness
of the local decision vector $\mathbf{x}_{t-\tau_{\text{l}}}^{c}$ is not useful at the master node.} \romannum{5}) construct gradient $\nabla\hat{f}_{t-\tau}^c(\hat{\mathbf{x}}_{t}^{c,j-1})$ in (\ref{EQ:EGDmaster}) with $\hat{\mathbf{d}}_{t-\tau_{\text{r}}}^c$ and $\hat{\mathbf{d}}_{t-\tau_{\text{r}}}^l$ replaced by $\hat{\mathbf{d}}_{t-\tau}^c$ and $\hat{\mathbf{d}}_{t-\tau}^l$, respectively, in Step 7; and \romannum{6}) solve \textbf{P2} with gradient $\nabla\hat{f}_{t-\tau}^c(\hat{\mathbf{x}}_{t}^{c,j-1})$ instead of $\nabla\hat{f}_{t-\tau_{\text{r}}}^c(\hat{\mathbf{x}}_{t}^{c,j-1})$ in Step 8. 

We make the following changes to Algorithm \ref{Alg:Worker}: \romannum{1}) start the algorithm at $t>\tau$ in Step 2; \romannum{2}) change $\hat{\mathbf{d}}_{t-\tau_{\text{r}}}^l$ to $\hat{\mathbf{d}}_{t-\tau}^l$ in Step~3; \romannum{3}) construct gradient $\nabla\hat{f}_{t-\tau_{\text{l}}}^c(\tilde{\mathbf{x}}_{t}^{c,j-1})$ in (\ref{EQ:EGDworkerc}) with $\mathbf{d}_t^c$ and $\hat{\mathbf{d}}_{t-\tau_\text{r}}^l$ replaced by $\mathbf{d}_{t-\tau_\text{l}}^c$ and $\hat{\mathbf{d}}_{t-\tau}^l$, respectively, in Step 6; \romannum{5}) solve \textbf{P3} with gradient $\nabla\hat{f}_{t-\tau_{\text{l}}}^c(\tilde{\mathbf{x}}_{t}^{c,j-1})$ instead of $\nabla\hat{f}_t^c(\tilde{\mathbf{x}}_{t}^{c,j-1})$ in Step 7; and \romannum{4}) modify $\mathbf{d}_t^c$ to $\mathbf{d}_{t-\tau_{\text{l}}}^c$ in Step~10.

Using similar techniques as those in the proof of Theorem \ref{thm}, we provide dynamic regret bounds for HiOCO in the presence of both local and remote delay.

\begin{theorem}\label{thm:LDL}
For any $\alpha\ge{L}$, $\xi>0$ and $\gamma\in(0,2\mu)$, the dynamic regret
yielded by HiOCO is bounded as follows:
\begin{enumerate}
\item[\romannum{1})] For any  $J_{\text{l}}+J_{\text{r}}\ge1$, the bound in claim \romannum{1}) of Theorem~\ref{thm} still holds by replacing $\tau_{\text{r}}$ with $\tau$.
\item[\romannum{2})] For any $J_{\text{l}}+J_{\text{r}}\ge1$ such that $4\eta^{J_{\text{l}}+J_{\text{r}}}<1$,
we have
\begin{align*}
        \text{RE}_T^{\text{d}}&\le\frac{1}{2\xi}\sum_{t=1}^T\Vert\nabla{f}_t(\mathbf{x}_t^\star)\Vert^2+\frac{L+\xi}{2}\tau{R}^2+\frac{L+\xi}{2(1-4\eta^{J_{\text{l}}+J_{\text{r}}})}\left(\!\tau{R}^2+6\tau^2\Pi_{2,T}^\star+\frac{4\beta}{1-\eta}\Delta_{2,T}\!\right)\!.
\end{align*}
\end{enumerate}
\end{theorem}
\textit{Proof:} See Appendix \ref{APP:localdelay}.

Due to the additional local delay, Theorem \ref{thm:LDL} has a more stringent condition on the total number of gradient descent steps in claim \romannum{2}) compared with the one in Theorem \ref{thm}. However, the order of the dynamic regret bound is still dominated by the accumulated system variation measures and is the same as the case without local delay. In particular, Theorem \ref{thm:LDL} implies that Corollary \ref{Cor-BD} still holds for the case of non-zero local delay by replacing $\tau_{\text{r}}$ with $\tau$.

\section{Conclusions}
\label{Sec:Conclusions}

We have studied a new OCO framework over a master-worker network, where the local data at the worker nodes may be non-independent or non-identically distributed, the global cost functions may be non-separable, and there may be multi-slot delay in both local data acquisition and the communication between the worker nodes and the master node. We propose the HiOCO algorithm, which takes full advantage of the network heterogeneity in information timeliness and computation capacity. HiOCO allows both timely local gradient descent at the local worker nodes and delayed global gradient descent at the remote master node. Furthermore, HiOCO allows multi-step estimated gradient descent at both the worker nodes and the master node to fully utilize their computation capacities. Our analysis considers the impacts of the unique hierarchical architecture, multi-slot delay, and gradient estimation error, on the performance guarantees of HiOCO in terms of dynamic regret bounds.
\appendices

\section{Proof of Lemma \ref{lm:GDE}}\label{app:GDE}
\textit{Proof:} Note that $\langle\nabla\hat{f}(\mathbf{y}),\mathbf{x}-\mathbf{y}\rangle+\frac{\alpha}{2}\Vert\mathbf{x}-\mathbf{y}\Vert^2$
 is $\alpha$-strongly convex. Applying Lemma \ref{lm:StrConv}, we have
 \small
\begin{align}
        &\langle\nabla{f}(\mathbf{y}),\mathbf{z}-\mathbf{y}\rangle+\frac{\alpha}{2}\Vert\mathbf{z}-\mathbf{y}\Vert^2+\langle\nabla\hat{f}(\mathbf{y})-\nabla{f}(\mathbf{y}),\mathbf{z}-\mathbf{y}\rangle\notag\\
        &\qquad\le\langle\nabla{f}(\mathbf{y}),\mathbf{x}^{\star}-\mathbf{y}\rangle+\frac{\alpha}{2}\Vert\mathbf{x}^\star-\mathbf{y}\Vert^2-\frac{\alpha}{2}\Vert\mathbf{z}-\mathbf{x}^\star\Vert^2+\langle\nabla\hat{f}(\mathbf{y})-\nabla{f}(\mathbf{y}),\mathbf{x}^\star-\mathbf{y}\rangle.\label{eq:lmGDE-1}
\end{align}
\normalsize
From $f(\mathbf{x})$ being $L$-smooth, we have
\small
\begin{align}
        f(\mathbf{z})\le{f}(\mathbf{y})+\langle\nabla{f}(\mathbf{y}),\mathbf{z}-\mathbf{y}\rangle+\frac{L}{2}\Vert\mathbf{z}-\mathbf{y}\Vert^2.\label{eq:lmGDE-2}
\end{align}
\normalsize
From $f_{t}(\mathbf{x})$ being $\mu$-strongly convex,
we have
\small
\begin{align}
        f(\mathbf{x}^\star)\ge{f}(\mathbf{y})+\langle\nabla{f}(\mathbf{y}),\mathbf{x}^{\star}-\mathbf{y}\rangle+\frac{\mu}{2}\Vert\mathbf{x}^{\star}-\mathbf{y}\Vert^2.\label{eq:lmGDE-3}
\end{align}
\normalsize
Adding $f(\mathbf{y})$ on both sides of (\ref{eq:lmGDE-1}), and then applying
(\ref{eq:lmGDE-2})
and (\ref{eq:lmGDE-3}) to the LHS and RHS of (\ref{eq:lmGDE-1}), respectively,
we have
\small
\begin{align}
        &{f}(\mathbf{z})-\frac{L}{2}\Vert\mathbf{z}-\mathbf{y}\Vert^2+\frac{\alpha}{2}\Vert\mathbf{z}-\mathbf{y}\Vert^2\notag\\
        &\qquad\le{f}(\mathbf{x}^\star)-\frac{\mu}{2}\Vert\mathbf{y}-\mathbf{x}^\star\Vert^2+\frac{\alpha}{2}\Vert\mathbf{y}-\mathbf{x}^{\star}\Vert^2-\frac{\alpha}{2}\Vert\mathbf{z}-\mathbf{x}^\star\Vert^2+\langle\nabla{f}(\mathbf{y})-\nabla\hat{f}(\mathbf{y}),\mathbf{z}-\mathbf{x}^{\star}\rangle.\label{eq:lmGDE-4}
\end{align}
\normalsize
Applying Lemma \ref{lm:StrConv} again, we have
\small
\begin{align}
        f(\mathbf{x}^\star)\le{f}(\mathbf{z})-\frac{\mu}{2}\Vert\mathbf{z}-\mathbf{x}^\star\Vert^2.\label{eq:lmGDE-5}
\end{align}
\normalsize

Substituting (\ref{eq:lmGDE-5}) into the RHS of (\ref{eq:lmGDE-4}),  noting
that $\langle\mathbf{a},\mathbf{b}\rangle\le\frac{1}{2\gamma}\Vert\mathbf{a}\Vert^2+\frac{\gamma}{2}\Vert\mathbf{b}\Vert^2$
for any $\gamma>0$, multiplying both sides by $2$, and rearranging terms, we have 
\small
\begin{align}
        \left(\alpha+\mu-\gamma\right)\Vert\mathbf{z}-\mathbf{x}^\star\Vert^2+\left(\alpha-L\right)\Vert\mathbf{z}-\mathbf{y}\Vert^2\le\left(\alpha-\mu\right)\Vert\mathbf{y}-\mathbf{x}^{\star}\Vert^2+\frac{1}{\gamma}\Vert\nabla{f}(\mathbf{y})-\nabla\hat{f}(\mathbf{y})\Vert^2.\label{eq:lmGDE-6}
\end{align}
\normalsize

Note that the strong convexity constant $\mu$ is smaller than the constant of smoothness $L$, \ie $\mu\le{L}$ \cite{Mokhtari16}. From (\ref{eq:lmGDE-6}), we have (\ref{eq:GDE-1}) for any $\alpha\ge{L}$ and $\gamma<2\mu\le\alpha+\mu$.\endIEEEproof

\section{Proof of Theorem \ref{thm}}\label{app:thm}
\textit{Proof:} We first prove claim \romannum{1}). We have
\small
\begin{align}
        \text{RE}_T^{\text{d}}&=\sum_{t=1}^T\left(f_t(\mathbf{x}_t)-f_t(\mathbf{x}_t^\star)\right)\stackrel{(a)}{\le}\sum_{t=1}^T\langle\nabla{f}_t(\mathbf{x}_t),\mathbf{x}_t-\mathbf{x}_t^\star\rangle\notag\\
        &\stackrel{(b)}{\le}\sum_{t=1}^T\Vert\nabla{f}_t(\mathbf{x}_t)\Vert\Vert\mathbf{x}_t-\mathbf{x}_t^\star\Vert\stackrel{(c)}{\le}\tau_{\text{r}}{DR}+D\sum_{t=\tau_{\text{r}}+1}^T\Vert\mathbf{x}_t-\mathbf{x}_t^\star\Vert.\label{EQ:thm-4}
\end{align}
\normalsize
where $(a)$ follows from the convexity of $f_t(\mathbf{x})$; $(b)$ is because
$\langle\mathbf{a},\mathbf{b}\rangle\le\Vert\mathbf{a}\Vert\Vert\mathbf{b}\Vert$;
and $(c)$ follows from $\nabla{f}_t(\mathbf{x})$ and $\mathcal{X}$ being
bounded in (\ref{EQ:D}) and (\ref{EQ:R}), respectively.

We now bound $\sum_{t=\tau_{\text{r}}+1}^T\Vert\mathbf{x}_t-\mathbf{x}_t^\star\Vert$
in (\ref{EQ:thm-4}). We have
\small
\begin{align}
        \sum_{t=\tau_{\text{r}}+1}^T\Vert\mathbf{x}_t-\mathbf{x}_t^\star\Vert\stackrel{(a)}{\le}\sqrt{\eta^{J_{\text{l}}}}\sum_{t=\tau_{\text{r}}+1}^T\left(\Vert\hat{\mathbf{x}}_t^{J_{\text{r}}}-\mathbf{x}_t^\star\Vert\right)+\frac{1-\sqrt{\eta^{J_{\text{l}}}}}{1-\sqrt{\eta}}\sqrt{\beta}\Delta_T\label{EQ:thm-5-1}
\end{align}
\normalsize
where  $(a)$ follows from applying Lemma~\ref{lm:GDE} to \textbf{P3} for $J_{\text{l}}$ times, $\Vert\mathbf{a}\Vert^2+\Vert\mathbf{b}\Vert^2\le(\Vert\mathbf{a}\Vert+\Vert\mathbf{b}\Vert)^2$
such that $\Vert\tilde{\mathbf{x}}_t^{j}-\mathbf{x}_t^\star\Vert\le\sqrt{\eta}\Vert\tilde{\mathbf{x}}_t^{j-1}-\mathbf{x}_t^\star\Vert+\sqrt{\beta}\Vert\nabla{f}_t(\tilde{\mathbf{x}}_t^{j-1})-\nabla\hat{f}_t(\tilde{\mathbf{x}}_t^{j-1})\Vert$
for any $j\in\{1,\dots,J_{\text{l}}\}$, and the definition of $\Delta_T$
in (\ref{EQ:DeltaT}). We continue to bound $\sum_{t=\tau_{\text{r}}+1}^T\Vert\hat{\mathbf{x}}_t^{J_{\text{r}}}-\mathbf{x}_t^\star\Vert$
in (\ref{EQ:thm-5-1}) as follows:
\small
\begin{align}
        \sum_{t=\tau_{\text{r}}+1}^T\Vert\hat{\mathbf{x}}_t^{J_{\text{r}}}-\mathbf{x}_t^\star\Vert&\stackrel{(a)}{\le}\sum_{t=\tau_{\text{r}}+1}^T\left(\Vert\hat{\mathbf{x}}_t^{J_{\text{r}}}-\mathbf{x}_{t-\tau_{\text{r}}}^\star\Vert+\Vert\mathbf{x}_t^{\star}-\mathbf{x}_{t-\tau_{\text{r}}}^\star\Vert\right)\stackrel{(b)}{\le}\sum_{t=\tau_{\text{r}}+1}^T\left(\Vert\hat{\mathbf{x}}_t^{J_{\text{r}}}-\mathbf{x}_{t-\tau_{\text{r}}}^\star\Vert\right)+\tau_{\text{r}}\Pi_T^\star\notag\\
        &\stackrel{(c)}{\le}\sqrt{\eta^{J_{\text{r}}}}\sum_{t=\tau_{\text{r}}+1}^T\left(\Vert\mathbf{x}_{t-\tau_{\text{r}}}-\mathbf{x}_{t-\tau_{\text{r}}}^\star\Vert\right)\!+\!\tau_{\text{r}}\Pi_T^\star+\frac{1-\sqrt{\eta^{J_{\text{r}}}}}{1-\sqrt{\eta}}\sqrt{\beta}\Delta_T\label{EQ:thm-5}
\end{align}
where $(a)$ is because $\Vert\mathbf{a}+\mathbf{b}\Vert\le\Vert\mathbf{a}\Vert+\Vert\mathbf{b}\Vert$,
$(b)$ follows from the definition of $\Pi_T^\star$ in (\ref{EQ:PiT}), and
$(c)$ follows from applying Lemma~\ref{lm:GDE} to \textbf{P2} for $J_{\text{r}}$ times similar to $(a)$ in (\ref{EQ:thm-5-1}).

Substituting (\ref{EQ:thm-5}) into (\ref{EQ:thm-5-1}), noting that $\sum_{t=\tau_{\text{r}}+1}^{T}\Vert\mathbf{x}_{t-\tau_{\text{r}}}-\mathbf{x}_{t-\tau_{\text{r}}}^\star\Vert\le\sum_{t=1}^T\Vert\mathbf{x}_t-\mathbf{x}_t^\star\Vert$,
and rearranging terms,
we have
\small
\begin{align}
        \!\!\left(1\!-\!\sqrt{\eta^{J_{\text{l}}+J_{\text{r}}}}\right)\sum_{t=\tau_{\text{r}}+1}^T\Vert\mathbf{x}_t-\mathbf{x}_t^\star\Vert\!-\!\sqrt{\eta^{J_{\text{l}}+J_{\text{r}}}}\sum_{t=1}^{\tau_{\text{r}}}\Vert\mathbf{x}_t-\mathbf{x}_t^\star\Vert\!\le\!\sqrt{\eta^{J_{\text{l}}}}\tau_{\text{r}}\Pi_T^\star\!+\!\frac{\sqrt{\eta^{J_{\text{l}}}}(1\!-\!\sqrt{\eta^{J_{\text{r}}}})+1\!-\!\sqrt{\eta^{J_{\text{l}}}}}{1-\eta}\sqrt{\beta}\Delta_T.\!\!\label{EQ:thm-6}
\end{align}
\normalsize
Substituting (\ref{EQ:thm-6}) into (\ref{EQ:thm-4}) and noting that $\eta<1$ and the radius of $\mathcal{X}$ being bounded in (\ref{EQ:R}),
we prove claim \romannum{1}). 

We now prove claim \romannum{2}). We have
\small
\begin{align}
        \text{RE}_T^{\text{d}}&=\sum_{t=1}^T\left(f_t(\mathbf{x}_t)-f_t(\mathbf{x}_t^\star)\right)\stackrel{(a)}{\le}\sum_{t=1}^T\left(\langle\nabla{f}_t(\mathbf{x}_t^\star),\mathbf{x}_t-\mathbf{x}_t^\star\rangle+\frac{L}{2}\Vert\mathbf{x}_t-\mathbf{x}_t^\star\Vert^2\right)\notag\\
        &\stackrel{(b)}{\le}\frac{1}{2\xi}\sum_{t=1}^T\Vert\nabla{f}_t(\mathbf{x}_t^\star)\Vert^2+\frac{L\!+\!\xi}{2}\sum_{t=1}^T\Vert\mathbf{x}_t-\mathbf{x}_{t}^\star\Vert^2\stackrel{(c)}\le\!\frac{1}{2\xi}\!\sum_{t=1}^T\Vert\nabla{f}_t(\mathbf{x}_t^\star)\Vert^2\!+\!\frac{L\!+\!\xi}{2}\left(\!\tau_{\text{r}}{R}^2+\!\!\sum_{t=\tau_{\text{r}}+1}^T\!\!\Vert\mathbf{x}_t\!-\!\mathbf{x}_{t}^\star\Vert^2\!\right)\!\!\!\label{EQ:thm-1}
\end{align}
\normalsize
where $(a)$ follows from $f_t(\mathbf{x})$ being $L$-smooth in (\ref{EQ:Smooth}),
$(b)$ is because
$\langle\mathbf{a},\mathbf{b}\rangle\le\frac{1}{2\xi}\Vert\mathbf{a}\Vert^2+\frac{\xi}{2}\Vert\mathbf{b}\Vert^2$
for any $\xi>0$, and $(c)$ follows from the radius of $\mathcal{X}$ being
bounded in (\ref{EQ:R}).

We now bound $\sum_{t=\tau_{\text{r}}+1}^T\Vert\mathbf{x}_t-\mathbf{x}_t^\star\Vert^2$
in (\ref{EQ:thm-1}). We have
\small
\begin{align}
        \sum_{t=\tau_{\text{r}}+1}^T\Vert\mathbf{x}_t-\mathbf{x}_t^\star\Vert^2\stackrel{(a)}{\le}\sum_{t=\tau_{\text{r}}+1}^T\left(\eta^{J_{\text{l}}}\Vert\hat{\mathbf{x}}_t^{J_{\text{r}}}-\mathbf{x}_t^\star\Vert^2\right)+\frac{1-\eta^{J_{\text{l}}}}{1-\eta}\beta\Delta_{2,T}\label{EQ:thm-2-1}
\end{align}
\normalsize
where $(a)$ follows from applying Lemma~\ref{lm:GDE} to \textbf{P3} for $J_{\text{l}}$
times such that $\Vert\mathbf{x}_t-\mathbf{x}_t^\star\Vert^2\le\eta^{J_{\text{l}}}\Vert\hat{\mathbf{x}}_t^{J_{\text{r}}}-\mathbf{x}_t^\star\Vert^2+\beta\sum_{i=1}^{J_{\text{l}}}\eta^{i-1}\Vert\nabla\hat{f}_t(\tilde{\mathbf{x}}_t^{J_{\text{l}}-i})-\nabla{f}_t(\tilde{\mathbf{x}}_t^{J_{\text{l}}-i})\Vert^2$
for any $t>\tau_{\text{r}}$, and the definition of $\Delta_{2,T}$ in (\ref{EQ:Delta2T}).
We continue to bound $\sum_{t=\tau_{\text{r}}+1}^T\Vert\hat{\mathbf{x}}_t^{J_{\text{r}}}-\mathbf{x}_t^\star\Vert^2$
in (\ref{EQ:thm-2-1}) as follows:
\small
\begin{align}
        \sum_{t=\tau_{\text{r}}+1}^T\Vert\hat{\mathbf{x}}_t^{J_{\text{r}}}-\mathbf{x}_t^\star\Vert^2&\stackrel{(a)}{\le}2\sum_{t=\tau_{\text{r}}+1}^T(\Vert\hat{\mathbf{x}}_t^{J_{\text{r}}}-\mathbf{x}_{t-\tau_{\text{r}}}^\star\Vert^2+\Vert\mathbf{x}_t^\star-\mathbf{x}_{t-\tau_{\text{r}}}^\star\Vert^2)\stackrel{(b)}{\le}2\sum_{t=\tau_{\text{r}}+1}^T\left(\Vert\hat{\mathbf{x}}_t^{J_{\text{r}}}-\mathbf{x}_{t-\tau_{\text{r}}}^\star\Vert^2\right)+2\tau_{\text{r}}^2\Pi_{2,T}^\star\notag\\
        &\stackrel{(c)}{\le}2\sum_{t=\tau_{\text{r}}+1}^{T}\left(\eta^{J_\text{r}}\Vert\mathbf{x}_{t-\tau_{\text{r}}}-\mathbf{x}_{t-\tau_{\text{r}}}^\star\Vert^2\right)+2\tau_{\text{r}}^2\Pi_{2,T}^\star+\frac{2(1-\eta^{J_\text{r}})}{1-\eta}\beta\Delta_{2,T}\label{EQ:thm-2}
\end{align}
\normalsize
where $(a)$ is because $\Vert\mathbf{a}+\mathbf{b}\Vert^2\le2(\Vert\mathbf{a}\Vert^2+\Vert\mathbf{b}\Vert^2)$,
$(b)$ follows from the definition of $\Pi_{2,T}^\star$ in (\ref{EQ:Pi2T})
and $|\sum_{i=1}^nx_i|\le\sum_{i=1}^n|x_i|\le\sqrt{n\sum_{i=1}^n|x_i|^2}$
such that $\Vert\mathbf{x}_{t}^{\star}-\mathbf{x}_{t-\tau_{\text{r}}}^\star\Vert^2\le\tau_{\text{r}}\sum_{i=1}^{\tau_{\text{r}}}\Vert\mathbf{x}_{t-\tau_{\text{r}}+i}^\star-\mathbf{x}_{t-\tau_{\text{r}}+i-1}^\star\Vert^2$,
and $(c)$ follows from applying Lemma~\ref{lm:GDE} to \textbf{P2} for $J_{\text{r}}$
times similar to $(a)$ in (\ref{EQ:thm-2-1}).

Substituting (\ref{EQ:thm-2}) into (\ref{EQ:thm-2-1}),  noting that $\sum_{t=\tau_{\text{r}}+1}^{T}\Vert\mathbf{x}_{t-\tau_{\text{r}}}-\mathbf{x}_{t-\tau_{\text{r}}}^\star\Vert^2\le\sum_{t=1}^T\Vert\mathbf{x}_t-\mathbf{x}_t^\star\Vert^2$,
and rearranging terms,
we have
\small
\begin{align}
        \left(1-2\eta^{J_{\text{l}}+J_{\text{r}}}\right)\sum_{t=\tau_\text{r}+1}^T\Vert\mathbf{x}_t-\mathbf{x}_t^\star\Vert^2-2\eta^{J_{\text{l}}+J_{\text{r}}}\sum_{t=1}^{\tau_{\text{r}}}\Vert\mathbf{x}_t-\mathbf{x}_t^\star\Vert^2\le2\eta^{J_{\text{l}}}\tau_{\text{r}}^2\Pi_{2,T}^\star+\frac{2\eta^{J_\text{l}}(1-\eta^{J_\text{r}})+1-\eta^{J_{\text{l}}}}{1-\eta}\beta\Delta_{2,T}.\label{EQ:thm-3}
\end{align}
\normalsize
Substituting into (\ref{EQ:thm-3}) into (\ref{EQ:thm-1}), noting that $\eta<1$
and the radius of $\mathcal{X}$ being bounded in (\ref{EQ:R}), and on the
condition $2\eta^{J_{\text{l}}+J_{\text{r}}}<1$, we complete the proof.
\endIEEEproof

\section{Proof of Theorem \ref{thm:LDL}}
\label{APP:localdelay}

\textit{Proof:} We first prove claim \romannum{1}). We can show that (\ref{EQ:thm-4}) still holds by replacing
$\tau_{\text{r}}$ with $\tau$. Applying Lemma \ref{lm:GDE} to \textbf{P3}
and \textbf{P2} for $J_{\text{l}}$ and $J_{\text{r}}$ times, respectively,
similar to the proofs of (\ref{EQ:thm-5-1}) and (\ref{EQ:thm-5}), we can show that
\small
\begin{align}
        &\sum_{t=\tau+1}^T\Vert\mathbf{x}_t-\mathbf{x}_t^\star\Vert\le\tau_{\text{l}}\Pi_T^\star+\sum_{t=\tau+1}^T\Vert\mathbf{x}_t-\mathbf{x}_{t-\tau_{\text{l}}}^\star\Vert\le\tau_{\text{l}}\Pi_T^\star+\sqrt{\eta^{J_{\text{l}}}}\sum_{t=\tau+1}^T\left(\Vert\hat{\mathbf{x}}_t^{J_{\text{r}}}-\mathbf{x}_{t-\tau_{\text{l}}}^\star\Vert\right)+\frac{1-\sqrt{\eta^{J_{\text{l}}}}}{1-\sqrt{\eta}}\sqrt{\beta}\Delta_T\notag\\
        &\le\tau_{\text{l}}\Pi_T^\star+\sqrt{\eta^{J_{\text{l}}}}\tau_{\text{r}}\Pi_T^\star+\sqrt{\eta^{J_{\text{l}}+J_{\text{r}}}}\sum_{t=\tau+1}^T\left(\Vert\mathbf{x}_{t-\tau}-\mathbf{x}_{t-\tau}^\star\Vert\right)+\frac{\sqrt{\eta^{J_{\text{l}}}}(1-\sqrt{\eta^{J_{\text{r}}}})}{1-\sqrt{\eta}}\sqrt{\beta}\Delta_T+\frac{1-\sqrt{\eta^{J_{\text{l}}}}}{1-\sqrt{\eta}}\sqrt{\beta}\Delta_T.\label{EQ:ThmL-3}
\end{align}     
\normalsize
Rearranging terms of (\ref{EQ:ThmL-3}) and then substituting it into the
version of (\ref{EQ:thm-4}) with $\tau_{\text{r}}$ replaced by $\tau$, we prove claim \romannum{1}).

We now prove claim \romannum{2}). We can show that (\ref{EQ:thm-1})
still holds by replacing $\tau_{\text{r}}$ with $\tau$ and 
\small
\begin{align}
       \sum_{t=\tau+1}^T\Vert\mathbf{x}_t-\mathbf{x}_{t}^\star\Vert^2\le2\sum_{t=\tau+1}^T(\Vert\mathbf{x}_t-\mathbf{x}_{t-\tau_{\text{l}}}^\star\Vert^2+\Vert\mathbf{x}_t^\star-\mathbf{x}_{t-\tau_{\text{l}}}^\star\Vert^2)\le2\tau_{\text{l}}^2\Pi_{2,T}^\star+2\sum_{t=\tau+1}^T\Vert\mathbf{x}_t-\mathbf{x}_{t-\tau_{\text{l}}}^\star\Vert^2.\label{EQ:ThmL-1}
\end{align}
\normalsize
By applying Lemma \ref{lm:GDE} to \textbf{P3} and \textbf{P2} for $J_{\text{l}}$
and $J_{\text{r}}$ times, respectively, similar to the proof of (\ref{EQ:thm-2}),
we can show that
\small
\begin{align}
        &\sum_{t=\tau+1}^T\Vert\mathbf{x}_t-\mathbf{x}_{t-\tau_{\text{l}}}^\star\Vert^2\le\eta^{J_{\text{l}}}\sum_{t=\tau+1}^T\left(\Vert\hat{\mathbf{x}}_t^{J_{\text{r}}}-\mathbf{x}_{t-\tau_{\text{l}}}^\star\Vert^2\right)+\frac{1-\eta^{J_{\text{l}}}}{1-\eta}\beta\Delta_{2,T}\notag\\
        &\qquad\le2\eta^{J_{\text{l}}}\!\!\sum_{t=\tau+1}^T\!\left(\Vert\hat{\mathbf{x}}_t^{J_{\text{r}}}-\mathbf{x}_{t-\tau}^\star\Vert^2\right)\!+2\eta^{J_{\text{l}}}\tau_{\text{r}}^2\Pi_{2,T}^\star+\frac{1-\eta^{J_{\text{l}}}}{1-\eta}\beta\Delta_{2,T}\notag\\
        &\qquad\le2\eta^{J_{\text{l}}+J_{\text{r}}}\sum_{t=\tau+1}^T(\Vert\mathbf{x}_{t-\tau}-\mathbf{x}_{t-\tau}^\star\Vert^2)+2\eta^{J_{\text{l}}}\frac{(1-\eta^{J_\text{r}})}{1-\eta}\beta\Delta_{2,T}+2\eta^{J_{\text{l}}}\tau_{\text{r}}^2\Pi_{2,T}^\star+\frac{1-\eta^{J_{\text{l}}}}{1-\eta}\beta\Delta_{2,T}.\label{EQ:ThmL-2}
\end{align}
\normalsize
Substituting (\ref{EQ:ThmL-2}) into (\ref{EQ:ThmL-1}) and rearranging terms,
we have
\small
\begin{align*}
        (1-4\eta^{J_{\text{l}}+J_{\text{r}}})\sum_{t=\tau+1}^T\Vert\mathbf{x}_t-\mathbf{x}_{t}^\star\Vert^2-4\eta^{J_{\text{l}}+J_{\text{r}}}\sum_{t=1}^{\tau}\Vert\mathbf{x}_t-\mathbf{x}_t^\star\le(2\tau_{\text{l}}^2+4\eta^{J_{\text{l}}}\tau_{\text{r}}^2)\Pi_{2,T}^\star\!+\!\left(\!4\eta^{J_{\text{l}}}\frac{(1\!-\!\eta^{J_\text{r}})}{1-\eta}+2\frac{1\!-\!\eta^{J_{\text{l}}}}{1-\eta}\!\right)\!\beta\Delta_{2,T}.
\end{align*}
\normalsize
Substituting the above inequality into the version of (\ref{EQ:thm-1}) with
$\tau_{\text{r}}$ replaced by $\tau$, noting that $\eta<1$, and on condition
that $4\eta^{J_{\text{l}}+J_{\text{r}}}<1$, we complete the proof. \endIEEEproof

\balance
\bibliographystyle{IEEEtran}
\bibliography{References}

\end{document}